\documentclass[%
 aip,
 amsmath,amssymb,
 reprint,%
]{revtex4-1}

\usepackage{graphicx}% Include figure files
\usepackage{dcolumn}% Align table columns on decimal point
\usepackage{bm}% bold math
\usepackage[utf8]{inputenc}
\usepackage[T1]{fontenc}
\usepackage{mathptmx}
\usepackage{etoolbox}

\makeatletter
\def\@email#1#2{%
 \endgroup
 \patchcmd{\titleblock@produce}
  {\frontmatter@RRAPformat}
  {\frontmatter@RRAPformat{\produce@RRAP{*#1\href{mailto:#2}{#2}}}\frontmatter@RRAPformat}
  {}{}
}%

\newcommand{\fref}[1]{Fig.~\ref{#1}}

\newcommand{\dd}{\mathrm{d}}

\def\mnras{Mon. Not. R. Astron. Soc.}

\def\apj{Astrophys. J.}
\def\apjl{Astrophys. J. Lett.}

\def\prd{Phys. Rev. D}
\def\prl{Phys. Rev. Lett.}

\def\aap{Astron. Astrophys.}

\def\jcap{J. Cosmology Astropart. Phys.}

\def\baas{Bull. Am. Astron. Soc.}
\def\nat{Nature}
\def\pasp{Publ. Astron. Soc. Pac.}

\begin{document}

\preprint{AIP/123-QED}

\title[Detecting intermediate-mass black hole binaries with atom interferometer observatories]{Detecting intermediate-mass black hole binaries with atom interferometer observatories: Using the resonant mode for the merger phase}

\author{Alejandro Torres-Orjuela}
 \email{atorreso@mail.sysu.edu.cn}
 \affiliation{MOE Key Laboratory of TianQin Mission, TianQin Research Center for Gravitational Physics \& School of Physics and Astronomy, Frontiers Science Center for TianQin, Gravitational Wave Research Center of CNSA, Sun Yat-Sen University (Zhuhai Campus), Zhuhai 519082, China}

\date{\today}% It is always \today, today,

\begin{abstract}
Atom interferometry detectors like AION, ZAIGA, and AEDGE will be able to detect gravitational waves (GWs) at dHz covering the band between large space-based laser interferometers LISA/TianQin/Taiji and ground-based facilities LIGO/Virgo/KAGRA. They will detect the late inspiral and merger of GW sources containing intermediate-mass black holes (IMBHs) in the mass range $10^2-10^5\,{\rm M_\odot}$. We study how accurately the parameters of an IMBH binary can be measured using the noise curve of AION. Furthermore, we propose a detection scheme where the early inspiral of the binary is detected using the regular broadband mode while the merger is detected using the resonant mode. We find that by using such a detection scheme the signal-to-noise ratio (SNR) of the detection as well as the detection accuracy of the parameters can be enhanced compared to the full detection of the signal using the broadband mode. We, further, assess the impact of the necessary detection gap while switching from broadband to resonant mode studying the case of a short ($30\,{\rm s}$) and a long ($600\,{\rm s}$) gap. We find that the improvement in SNR and detection accuracy is bigger for the shorter gap but that even in the case of the long gap such a scheme can be beneficial.
\end{abstract}

\maketitle

\section{Introduction}\label{sec:int}

With almost 100 confirmed gravitational wave (GW) detections~\cite{GWTC1,GWTC2,GWTC3}, GW astronomy is in full swing. Besides the existing laser interferometer observatories LIGO~\cite{ligo_2015}, Virgo~\cite{virgo_2012}, and KAGRA~\cite{kagra_2019} many other detectors based on laser interferometry have been proposed -- space-based observatories like LISA~\cite{lisa_2017}, TianQin~\cite{tianqin_2016}, Taiji~\cite{taiji_2015}, and DECIGO~\cite{decigo_2021} as well as future ground-based facilities such as the Einstein Telescope~\cite{et_2010} and Cosmic Explorer~\cite{cosmic_explorer_2019}. Besides laser interferometers, atom interferometers (AIs) have also been proposed to detect GWs~\cite{snadden_mcguirk_1998,dimopoulos_graham_2008}. Different systems for the demonstration of the technology have been established or are being built~\cite{zhou_xiong_2011,dickerson_hogan_2013,miga_2018,schlippert_meiners_2019,magis_2021} while ground-based observatories such as AION~\cite{aion_2020} and ZAIGA~\cite{zaiga_2020}, as well as space-based observatories like AEDGE~\cite{aedge_2020} are under planning.

AI will allow the study of multiple fields in GW astronomy including the detection of compact binaries~\cite{zhao_shao_2021,pujolas_vaskonen_2021,badurina_buchmueller_2022}, cosmology with standard sirens~\cite{cai_yang_2021,yang_lee_2022,yang_cai_2022}, studies of the stochastic GW background~\cite{croon_houtz_2019,ellis_lewicki_2020,ellis_lewicki_2021,campeti_komatsu_2021,barish_bird_2021,cui_sfakianakis_2021,chang_cui_2022}, and testing alternative theories of gravity~\cite{ellis_vaskonen_2020}. A remarkable feature of AI detectors is that they will cover the intermediate band in dHz between large space-based laser interferometers LISA/TianQin/Taiji and ground-based facilities LIGO/Virgo/KAGRA, therefore, being able to detect the late inspiral and merger of binaries in the intermediate-mass range $10^2-10^5\,{\rm M_\odot}$.

Intermediate-mass BHs (IMBHs) have been proposed as the ``seed'' BHs for the super-massive BHs in the center of most galaxies~\cite{volonteri_haardt_2003,volonteri_rees_2005}, to be lurking in the center of dwarf galaxies and globular clusters~\cite{barth_ho_2004,noyola_gebhardt_2008}, as dark matter candidates~\cite{carr_kuhnel_2016}, as the results of runaway collisions of massive stars in star clusters~\cite{portegies-zwart_mcmillan_2002,portegies-zwart_baumgardt_2004,gurkan_atakan_2004}, and as the engine powering miniquasars~\cite{madau_rees_2004}. Although playing a key role in several astrophysical systems, their detection has proved challenging and existing observations are often accompanied by big uncertainities~\cite{noyola_gebhardt_2010,webb_cseh_2012,lu_do_2013,lanzoni_mucciarelli_2013,kiziltan_baumgardt_2017,van-der-marel_anderson_2010,arca-sedda_2016,askar_arca-sedda_2018,weatherford_chatterjee_2018}. GW190521 represents the first direct detection of an IMBH forming after the merger of an $85\,{\rm M_\odot}$ BH with a $66\,{\rm M_\odot}$ BH~\cite{ligo_2020e}. However, this detection has provoked many discussions regarding the detection itself~\cite{romero-shaw_lasky_2020,fishbach_holz_2020,nitz_capano_2021,calderon-bustillo_sanchis-gual_2021b,olsen_roulet_2021,xu_hamilton_2022,gayathri_healy_2022,estelles_husa_2022} and its interpretation~\cite{chen_li_2019,calderon-bustillo_sanchis-gual_2021a,shibata_kiuchi_2021,torres-orjuela_chen_2022} because its short duration in band significantly limits the information contained. Therefore, the detection of IMBH binaries (IMBHB) and intermediate-mass ratio inspirals (IMRIs) emitting GWs in the dHz will be crucial in advancing our understanding of IMBHs~\cite{miller_2002,miller_hamilton_2002,fregeau_larson_2006,amaro-seoane_gair_2007,graff_buonanno_2015,amaro-seoane_2018b,arca-sedda_berry_2020,arca-sedda_amaro_seoane_2021}.

In this paper, we study the detection of an IMBHB merger using the noise curve of the AI detector AION-1km\cite{aion_2020} to analyze how accurately the parameters of the binary can be detected. In particular, we explore how the detection is impacted by using a scheme where the early inspiral is detected in the regular broadband mode while the merger is detected using the resonant mode. We evaluate how accurately the parameters of the IMBHB can be obtained from the early inspiral and if they allow a good prediction of the frequencies during the merger. We, further, study the impact of the necessary gap while switching from the broadband mode to the resonant mode assuming a short gap of $30\,{\rm s}$ and a long gap of $600\,{\rm s}$. Last, we compare the accuracy of the parameter extraction using the proposed scheme and the detection of the full signal in broadband mode.

\section{Gravitational wave detection with atom interferometer observatories}\label{sec:det}

Several AIs formed by cold atoms in free-fall that are operated simultaneously using a common laser source can be used to detect gravitational waves (GWs) via differential phase measurement~\cite{snadden_mcguirk_1998,dimopoulos_graham_2008,graham_hogan_2013,graham_hogan_2017,aion_2020}. The interferometry is performed in a vertical vacuum system, where the atom sources for the AIs are positioned along the length of the system. Laser pulses are used to drive transitions between the ground and excited states of the atoms, while also acting as beam splitters and mirrors for the atomic de Broglie waves. This way a quantum superposition of two paths -- a ground state and a long-lived ‘clock’ state -- is generated in each AI and later recombined. The phase imprinted along each path depends on the phase of the laser pulses that excite and de-excite the atoms as well as on the phase accumulated by the atoms themselves due to energy shifts. Therefore, the strain in the space between the free-falling atoms created by a passing GW induces a difference in the laser phase that can be measured using the spatially separated AIs.

AI detectors can, furthermore, be operated in the so-called resonant mode~\cite{graham_hogan_2016}. This operation mode can be accomplished using a series of $Q$ $\pi$-pulses in contrast to the broadband mode where only one $\pi$-pulse is used. Using the resonant mode has the advantage that we have a $Q$-fold enhancement of the detector's sensitivity. However, the drawback is that the detector becomes sensitive in only a small band with a width of $\sim f_r/Q$ around the resonance frequency $f_r := \pi/T$, where $2T$ is the interrogation time of the AI.

Despite the technical differences between AI and laser interferometer GW detectors, the data obtained from them can be analyzed in the same way using match filtering technics with the only major difference being that AI detectors are sensitive to GWs along one spatial direction. For the noise curve $S_n$ of the detector and two time-domain waveforms $h, h'$, we can define the usual noise-weighted inner product~\cite{finn_1992}
\begin{equation}\label{eq:innprod}
    \langle h, h'\rangle := 2\Re\left[\int_0^\infty\frac{\tilde{h}(f)\tilde{h'}(f)}{S_n(f)}\dd f\right]
\end{equation}
where $\tilde{h}$ and $\tilde{h'}$ are the Fourier transforms of $h$ and $h'$, respectively. The optimal signal-to-noise ratio (SNR) of a signal $h$ is then obtained when filtering with the same waveform
\begin{equation}\label{eq:snr}
    \rho^2 := \langle h, h\rangle.
\end{equation}

In the limit where the SNR is high, a Fisher matrix analysis~\citep{coe_2009} can be used to obtain linearized estimates for the measurement errors. Having the waveform of the GWs emitted by a source $h(\mathbf{\theta})$, where $\mathbf{\theta}$ are the intrinsic and extrinsic parameters of the source, the Fisher matrix is defined as
\begin{equation}\label{eq:deffm}
    \Gamma_{i,j} := \left\langle\frac{\partial h(\mathbf{\theta})}{\partial\theta_i},\frac{\partial h(\mathbf{\theta})}{\partial\theta_j}\right\rangle.
\end{equation}
The inverse of the Fisher matrix $C := \Gamma^{-1}$ approximates the sample covariance matrix of the Bayesian posterior distribution for the parameters given the observed signals. Therefore, the square root of a diagonal element $\sqrt{C_{ii}}$ gives the detection accuracy of the parameter $\mathbf{\theta}_i$ at the 1-$\sigma$ level while the non-diagonal elements $C_{ij}$ show the correlation between two different parameters $\mathbf{\theta}_i$ and $\mathbf{\theta}_j$~\cite{snecdecor_cochran_1991}.

\section{Combining the broadband and the resonant mode for detection}\label{sec:reso}

As mentioned in the previous section, AI GW detectors can be operated in a resonant mode that provides an enhanced sensitivity by a factor $Q$ but at the cost of only being sensitive in a smaller band with a width $\sim f_r/Q$ around a resonance frequency $f_r$. Using this operation mode, therefore, might be helpful when we want to detect an event where we know a priori its frequency range but, otherwise, it is more sensible to use the broadband mode to catch as many signals as possible.

\begin{figure}
\includegraphics[width=0.49\textwidth]{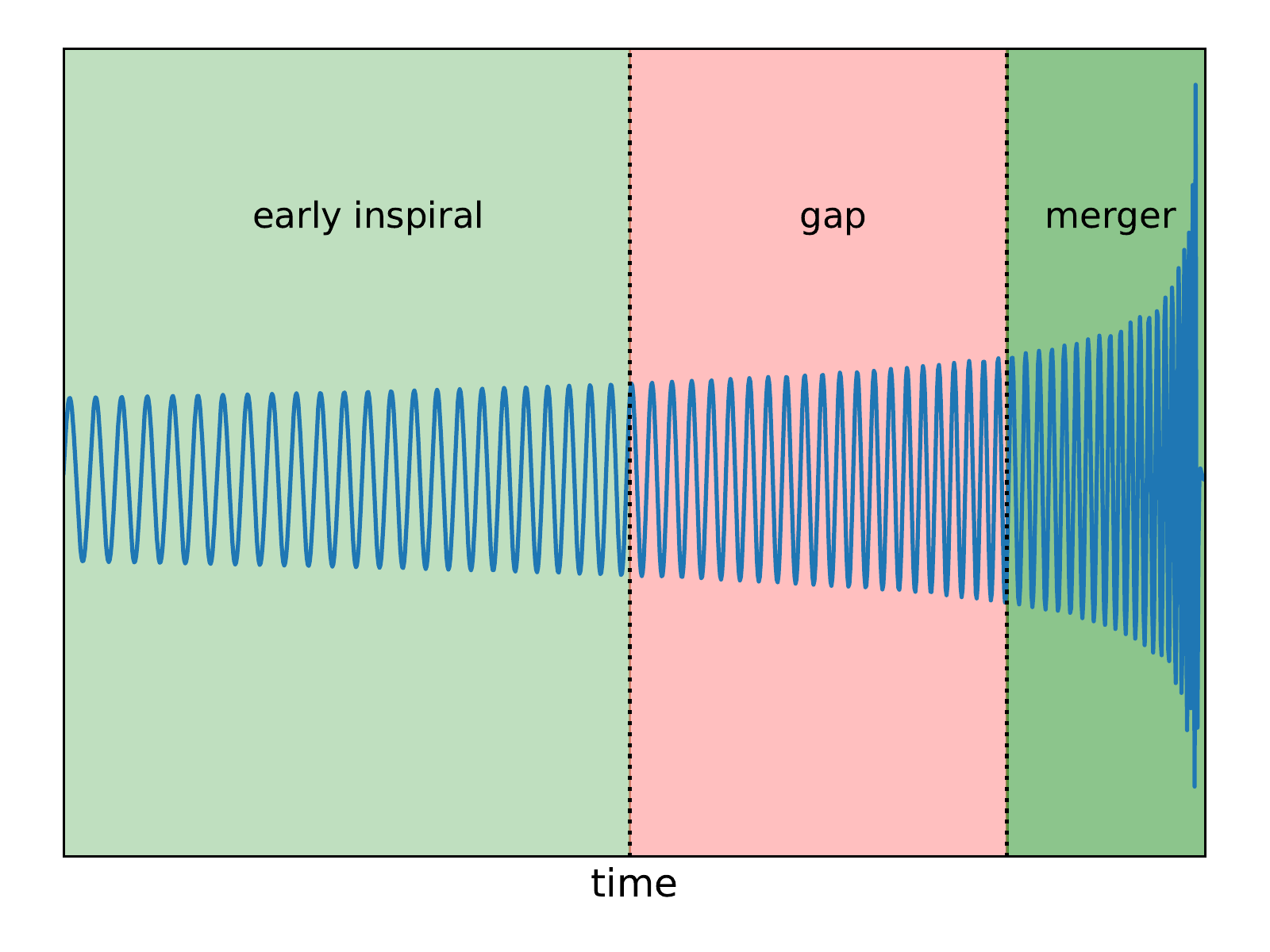}
    \caption{
        A scheme of the proposed detection scenario. The light green region shows the `early inspiral' used to estimate the parameters of the source to set the reference frequency and Q-factor. The red region is the `gap' in detection while the detector is being changed to its resonant mode. The dark green region shows the `merger' where the detector is operated in resonant mode.
    }\label{fig:scheme}
\end{figure}

Here we propose a detection scheme where both the broadband mode and the resonant mode are used to detect a single source. The detection scheme is shown in \fref{fig:scheme} and consists of detecting the `early inspiral' of the source in broadband mode, then a `gap' where the detector is switched to resonant mode and data is not recorded, and detecting the `merger' in resonant mode. Our goal is to explore if we can obtain an improved parameter estimation using such a detection scheme, where we study what impact different lengths of the gap have. Note that such a gap is necessary for two reasons: (i) switching the detector between its different modes might require some time and (ii) predicting the frequency band of the merger requires analyzing the data collected during the early inspiral.

Note that, in principle, analyzing the data recorded for the early inspiral can be done while the detector is still collecting data but, for simplicity, we assume that the data analysis and the change of operation mode are done during the same gap. Instead, we consider two gaps, `Gap 1' equal to $30\,{\rm s}$ and `Gap 2' equal to $600\,{\rm s}$, to explore the impact of a relatively short and a very long gap. In both cases, we assume the merger phase is detected from $45\,{\rm s}$ before the merger and that the early inspiral is recorded right until the gap starts. As an example, we consider the detection of an IMBHB with a total mass $M = 1400\,{\rm M_\odot}$ (in the detector frame), a mass ratio $q=3$, an inclination $\iota = \pi/4$, at a luminosity distance $D_L = 25\,{\rm Gpc}$. We further assume that the heavier BH has a spin aligned with the binary's angular momentum of magnitude $s_{1,z} = 0.7$ while the lighter BH has an aligned spin of magnitude $s_{2,z} = 0.5$ and that the wave has an initial phase $\phi_0 = 0$. We generate the waveform using the numerical relativity surrogate model \texttt{NRHybSur3dq8} that can accurately simulate the early inspiral and merger phase of binaries with mass ratios of up to 8 and aligned spins of up to 0.8~\cite{varma_field_2019b,field_galley_2014}. Moreover, we use the noise curve of AION-1km or short AION assuming the gravitational gradient noise can be fully modeled and mitigated~\cite{aion_2020}.

\subsection{Parameter estimation from the early inspiral}\label{ssec:inspiral}

\begin{figure}
\includegraphics[width=0.49\textwidth]{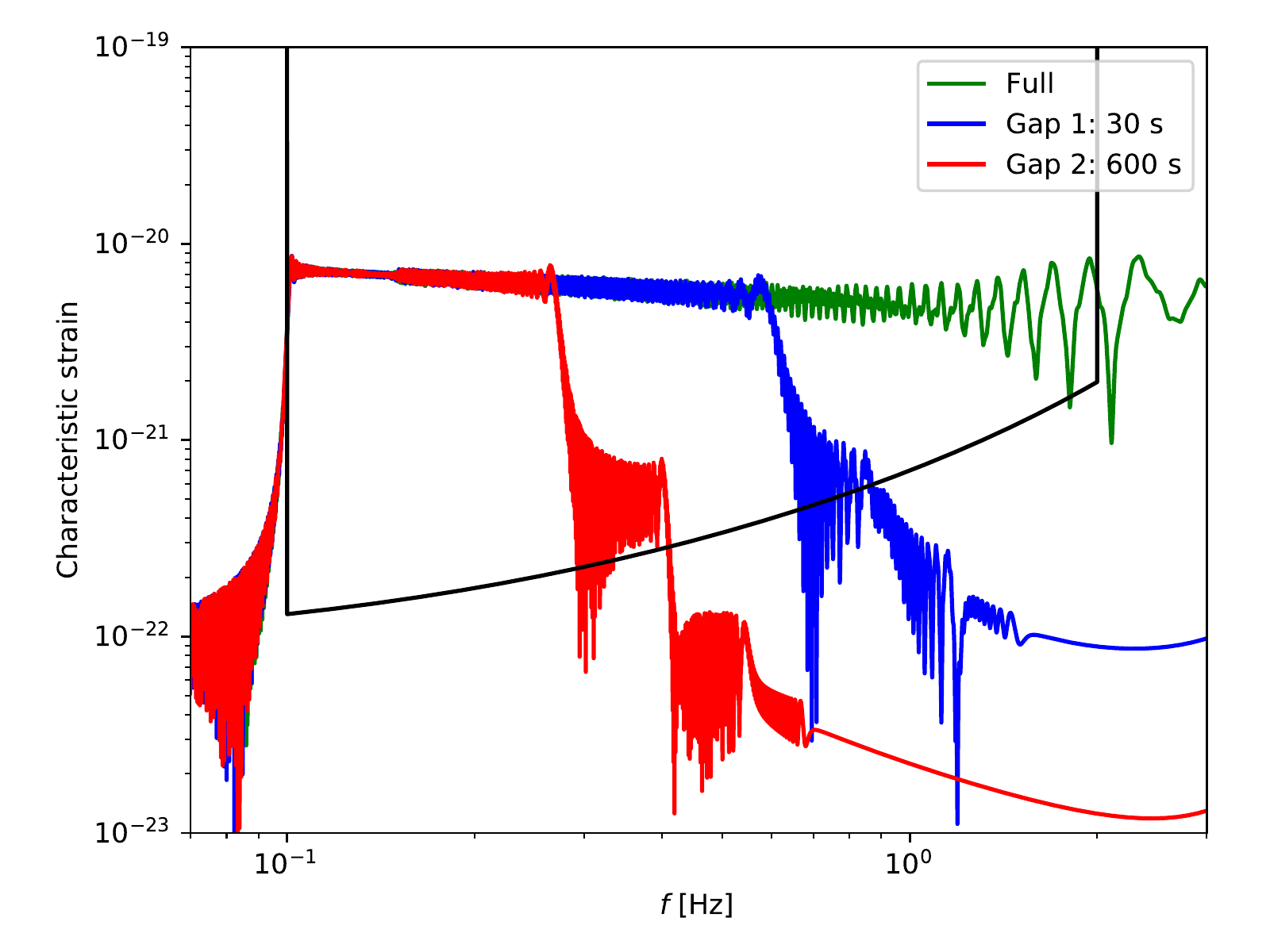}
    \caption{
        The characteristic strain for the full detection (green), for detection with a gap of $30\,{\rm s}$ (blue), and for detection with a gap of $600\,{\rm s}$ (red). The black line shows the PSD of AION.
    }\label{fig:strain}
\end{figure}

In the proposed detection scheme, the data of the early inspiral is used to predict the frequencies that we want to detect during the merger to set the reference frequency and the $Q$-factor for the resonant mode. Therefore, we study how accurately the parameters of the source can be detected using only the data before the gap performing a Fisher matrix analysis. In \fref{fig:strain}, we show the characteristic strain collected before the two gaps as well as for the full wave. We see that for the full detection, the characteristic strain remains at a similar level, well above the noise curve of AIONs, for the full width of the detection band. In contrast, the characteristic strain for `Gap1' ($30\,{\rm s}$) significantly decreases for frequencies above $6\,{\rm dHz}$ while the characteristic strain for `Gap 2' ($600\,{\rm s}$) falls off around $2.5\,{\rm dHz}$ and once again around $4\,{\rm dHz}$. Despite the decrease in the characteristic strain for higher frequencies when having gaps, we find that the SNR is not severely affected having a SNR $\rho_{\rm full} = 84.1$ for the full detection, $\rho_{\rm gap,1} = 83.3$ for `Gap1', and $\rho_{\rm gap,2} = 80.9$ for `Gap2'. The SNR only changes by a relatively small proportion when having gaps because the source spends $9475\,{\rm s}$ in the band, which is much longer than even the long gap.

\begin{figure*}
\includegraphics[width=0.98\textwidth]{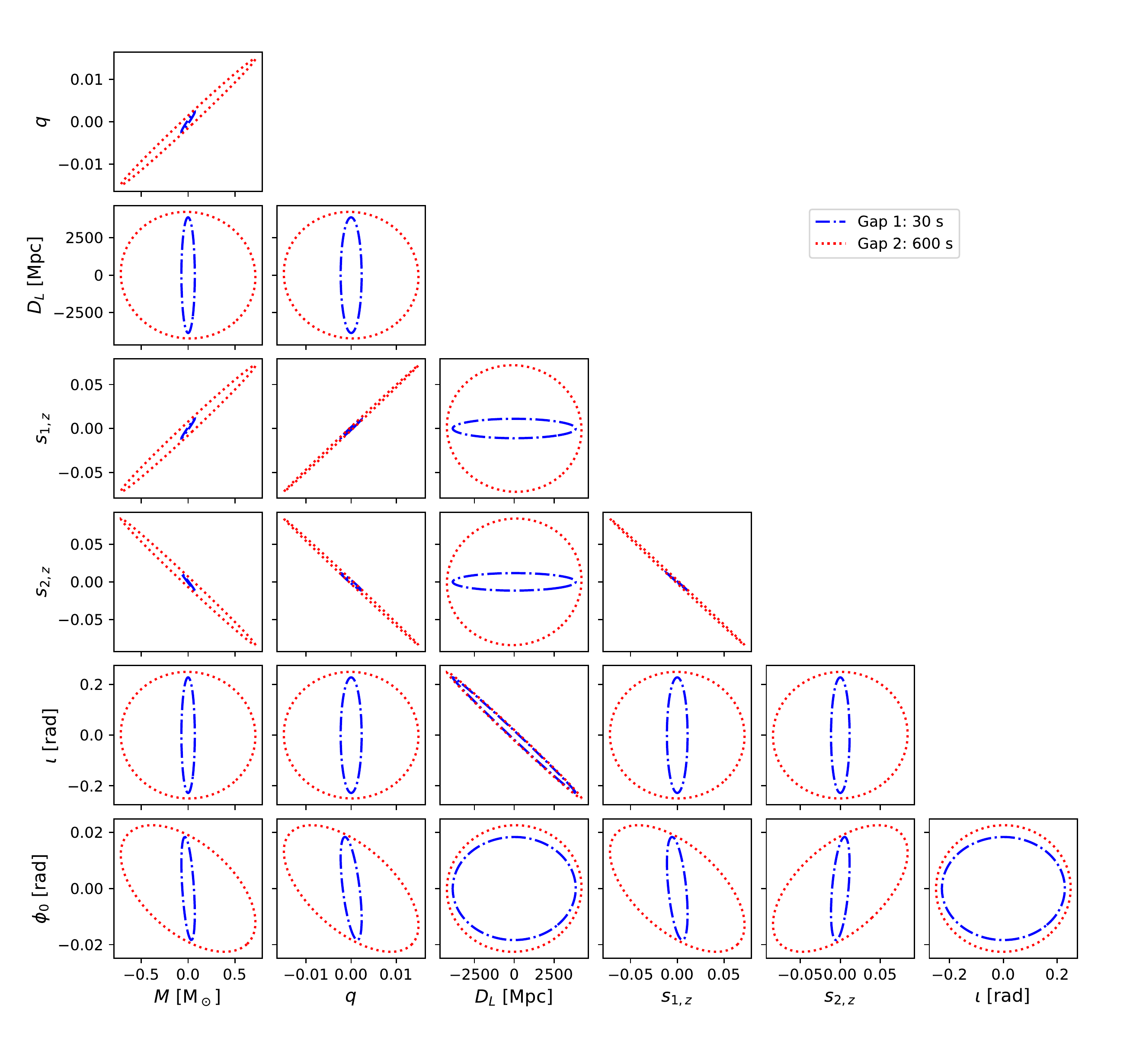}
    \caption{
        The covariance ellipses for the parameters of the IMBHB for detection of the early inspiral assuming a gap of $30\,{\rm s}$ (blue dashed-dotted line) and a gap of $600\,{\rm s}$ (red dotted line).
    }\label{fig:inspiral}
\end{figure*}

\fref{fig:inspiral} shows the detection accuracy for the seven parameters of the source that we need to set in the waveform model \texttt{NRHybSur3dq8}. As expected, the detection accuracy for the short gap of $30\,{\rm s}$ is better than for the long gap of $600\,{\rm s}$. The difference is particularly prominent for the parameters that are usually constrained the most during the late inspiral and the merger like the spins $s_{1,z}$ and $s_{2,z}$, and the mass ratio $q$. Note that the detection accuracy for the total mass is also significantly better for the short gap, however, in terms of relative error the total mass is constrained much better than all other parameters and thus the difference is not very significant. In general, we find that for both gaps the parameters are constrained with high accuracy having relative errors of $10\,\%$ or less.

\begin{figure}
\includegraphics[width=0.49\textwidth]{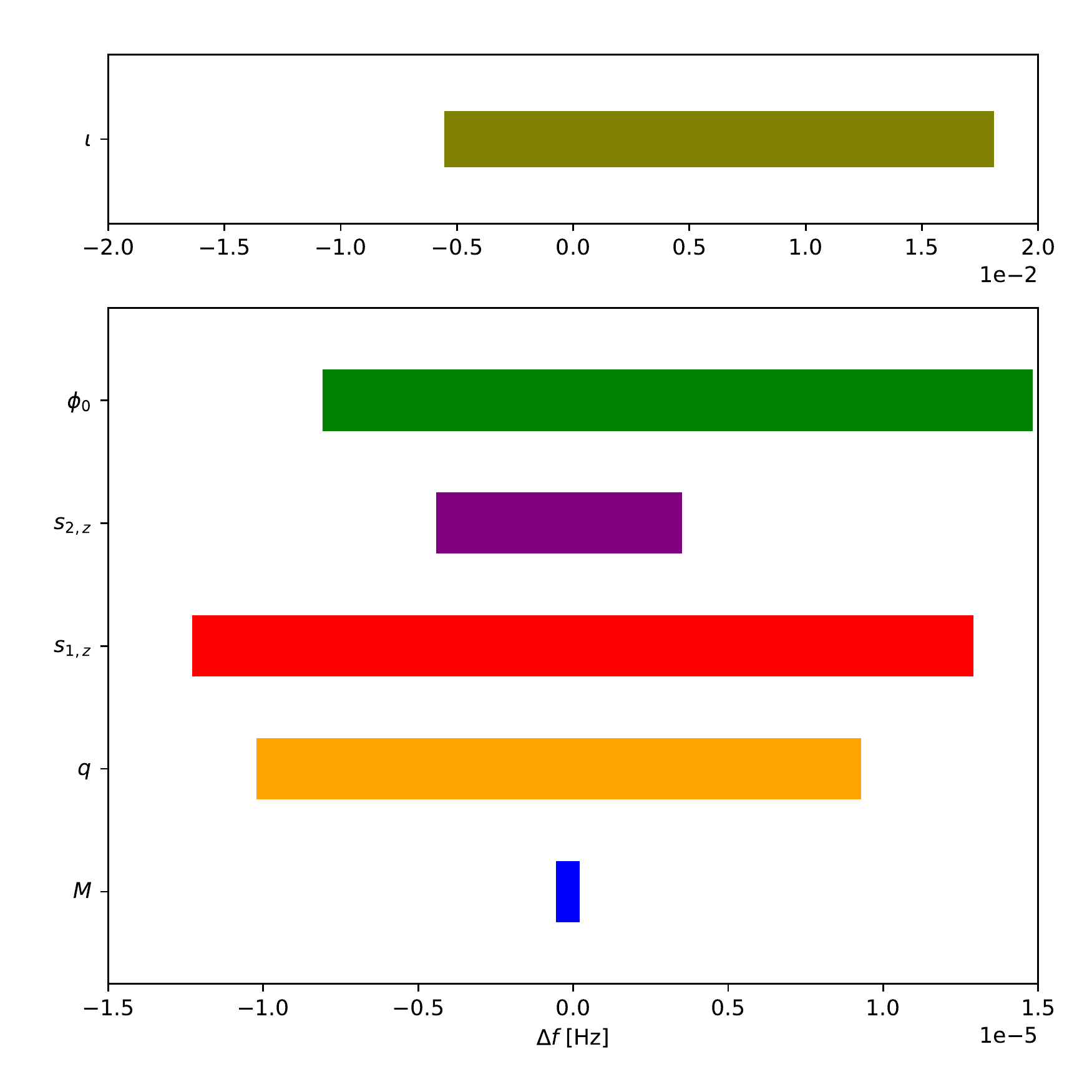}
    \caption{
        The error in the frequency at the merger for the errors of the parameters that affect the frequency. The errors in the parameters correspond to the errors from `Gap2'.
    }\label{fig:dfreq}
\end{figure}

The idea is to use the parameters obtained from the early inspiral to estimate the frequency range we want to detect around the merger. Therefore, a critical question is if their accuracy is good enough to obtain a good estimation of the frequencies. To evaluate how the error in the parameters impacts the frequencies we want to observe, we compare the frequency at merger we get when having the original parameters and when varying one of the parameters by the errors from `Gap2' (either adding or subtracting the error) shown in \fref{fig:inspiral}. We define the frequency at the merger as the maximal frequency in the band we get when using the function `frequency\_from\_polarizations' from \texttt{PyCBC}~\cite{pycbc_2020,pycbc_inference_2019} when considering the time interval $[-45,0]\,{\rm s}$ of the time domain waveform produced by \texttt{NRHybSur3dq8}. We see in \fref{fig:dfreq}, that the total mass $M$, the mass ratio $q$, the spins $s_{1,z}$ and $s_{2,z}$, as well as the initial phase $\phi_0$ only induce errors in the frequency at the order of $10^{-5}$. In contrast to what one would expect, the error in the total mass induces the smallest error in the frequency, although, $M$ dominates the evolution of the frequency~\cite{cutler_flanagan_1994}. However, we remember that the relative error of $M$ is smaller than for the other parameters and that error propagation is proportional to the relative error. The biggest error is induced by the error in the inclination $\iota$ which had to be shown separately since it is around three orders of magnitude bigger. However, the error is still at least one order of magnitude smaller than the frequencies considered $[1,20]\,{\rm dHz}$, and thus we do not expect a significant effect on the estimation of the frequency range. The luminosity distance is not considered because it only affects the amplitude of the wave and not its frequency, therefore, inducing a vanishing error.

We find that the errors in the parameters resulting from `Gap2' lead to very small errors that are one to four orders of magnitude smaller than the frequencies considered. For `Gap1' we do not show this dependency because the errors induced are even smaller, often below the waveform's numerical accuracy. Therefore, we proceed by assuming that we can obtain the original values of the parameters of the source and use them to estimate the frequency range we want to observe around the merger.

\subsection{Parameter estimation after the merger}\label{ssec:merger}

After determining the parameters of the source from the early inspiral before the gap, we can change the detector into its resonant mode to detect the merger with high accuracy. That means we want to have a $Q$-factor that is big enough to significantly enhance the detection accuracy. However, at the same time, $Q$ and the resonance frequency $f_r$ need to be set up so that the bandwidth ($\sim f_r/Q$) covers the evolution of the source's frequency during the late inspiral and the merger. In the previous subsection, we show that the parameter extraction from the early inspiral is accurate enough that it does not significantly limit our ability to predict the frequency range we want to detect. Thus we look for an ``optimal'' setup.

We determine the frequency range as follows: (i) take the last $45\,{\rm s}$ before the merger (including the ringdown) and Fourier transform the waveform, (ii) find the frequency corresponding to the highest amplitude $h_{\rm max}$ of the transformed wave, (iii) determine the interval $[f_{\rm low},f_{\rm high}]$ of frequencies where the amplitude of the wave is more than $0.75\times h_{\rm max}$, and (iv) set the resonance frequency to be the mean value of $[f_{\rm low},f_{\rm high}]$ and $Q$ so that $f_r/Q = f_{\rm high} - f_{\rm low}$. For the source considered, we get $[7.8,8.2]\,{\rm dHz}$ as the frequency range we want to detect, thus finding $f_r = 8.2\,{\rm dHz}$ and $Q = 10$.

Using AION in its resonant mode with the aforementioned parameters, we get $\rho_{\rm merger} = 24.3$ for the SNR of the source during the merger stage. Combining this SNR with the SNR accumulated before the gaps, where we use that $\rho_{\rm total,i}^2 = \rho_{\rm gap,i}^2 + \rho_{\rm merger}^2$, we get $\rho_{\rm total,1} = 87.3$ and $\rho_{\rm total,2} = 84.5$ for `Gap1' and `Gap2, respectively. Therefore, we find that for both gaps the SNR improves when compared to the SNR of the full detection when operating the detector all the time in broadband mode, where we have $\rho_{\rm full} = 84.1$.

A higher SNR usually allows a more accurate parameter extraction. Therefore, we compare the errors we get in the seven parameters of the source when having a full observation of the source but in broadband mode to when having the early inspiral in broadband mode, a gap to pre-analyze the source/change the detector, and the detection of the merger in resonant mode. We perform again a Fisher matrix analysis, where for the second scenario we use that $\Gamma^{\rm total}_{i,j} = \Gamma^{\rm inspiral}_{i,j} + \Gamma^{\rm merger}_{i,j}$ because the early inspiral and the merger happen at different frequencies~\cite{isoyama_nakano_2018}.

\begin{figure*}
\includegraphics[width=0.98\textwidth]{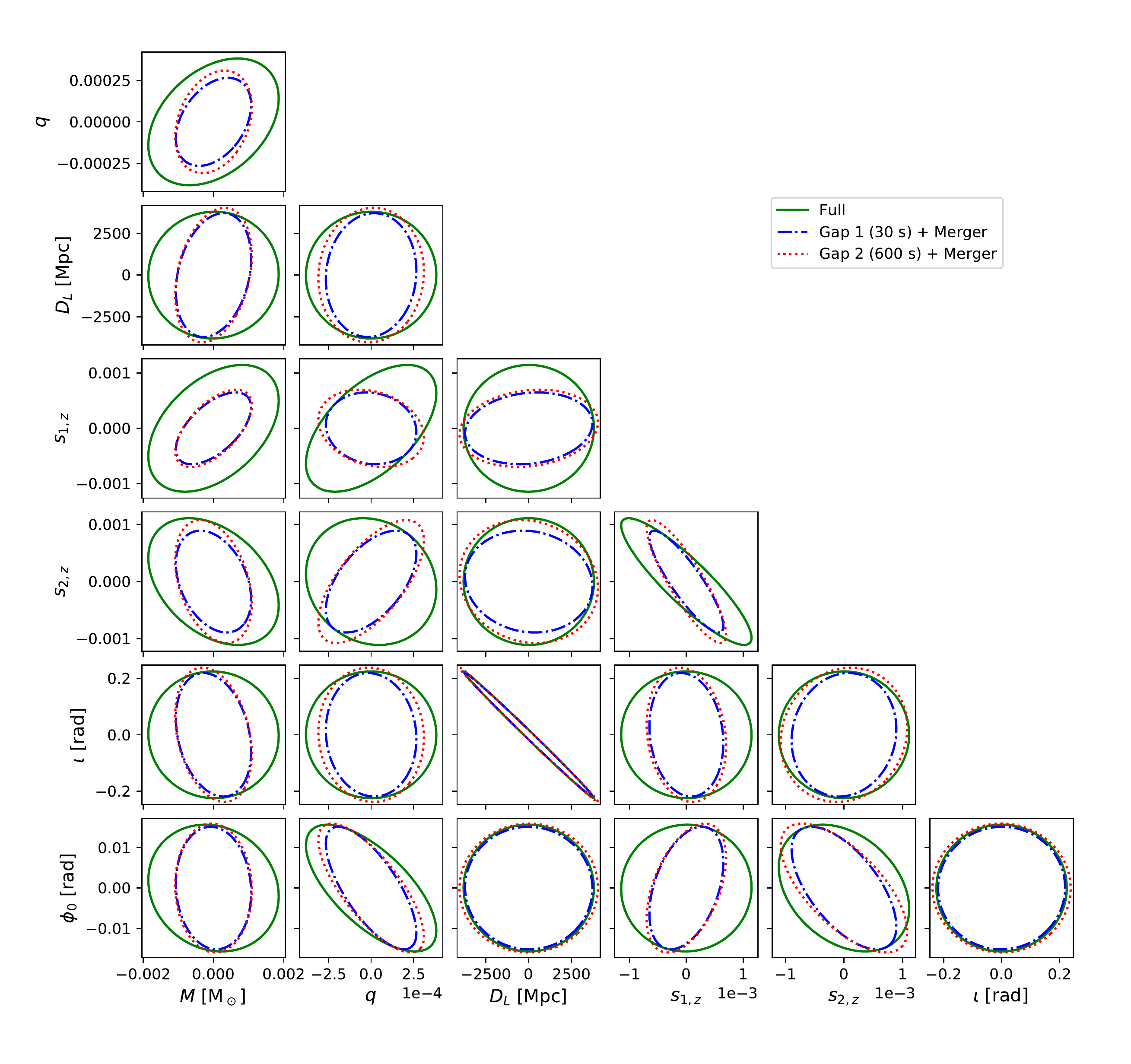}
    \caption{
        The covariance ellipses for the parameters of the IMBHB for detection of the full signal (green solid line), the early inspiral assuming a gap of $30\,{\rm s}$ plus the merger (blue dashed-dotted line), and assuming a gap of $600\,{\rm s}$ plus the merger (red dotted line).
    }\label{fig:merger}
\end{figure*}

In \fref{fig:merger}, we see that the detection accuracy for the early inspiral before `Gap1' ($30\,{\rm s}$) plus the merger in resonant mode is better than for the full wave detected in the broadband mode for all parameters. The improvement in detection accuracy is particularly prominent for parameters that are constrained the most during the late inspiral and merger like $q$, $s_{1,z}$, and $s_{2_z}$, but also the total mass $M$ is constrained more accurately. The remaining three parameters $D_L$, $\iota$, and $\phi_0$ are also tighter constrained than for the full detection but only marginally. In the case of the inspiral before `Gap2' ($600\,{\rm s}$) plus the merger, $D_L$, $\iota$, and $\phi_0$ are constrained a little worse than for the full detection while the spins $s_{1,z}$ and $s_{2,z}$ are constrained a little better. However, we find again that the total mass $M$ and the mass ratio $q$ are constrained significantly better than in the case of full detection. Besides getting a higher accuracy by combining the broadband and the resonant mode in detection, another interesting feature is that the (anti-)correlation between different parameters can be affected. In particular, we find that for the spins $s_{1,z}$ and $s_{2,z}$, and the mass ratio $q$ the relation changes from correlation and anti-correlation to anti-correlation and correlation, respectively, when changing from the full detection to the combined detection. Therefore, using the combined mode could help to disentangle parameters that are otherwise degenerate in GW detection.

\section{Conclusions}\label{sec:con}

In this paper, we study how a combination of the broadband mode and the resonant mode of AI observatories can be used to improve the detection of GW sources in the dHz band. We show that a detection scheme where the early inspiral is detected in the broadband mode, followed by a gap required to perform the parameter estimation and the change of the detector to resonant mode, completed by detecting the merger in the resonant mode can improve the detection accuracy compared to detecting the full signal in broadband mode. We explore the case of a long gap that lasts $600\,{\rm s}$ and a short gap that lasts only $30\,{\rm s}$. Although the duration of the short gap is optimistic, in particular considering the time required for detection and parameter estimation, a constant improvement in waveform modeling~\cite{field_galley_2014,varma_field_2019b,setyawati_ohme_2021,hamilton_london_2021,ramos-buades_buonanno_2023} combined with advanced technics such as neural networks, autoencoders, and machine learning~\cite{george_huerta_2016,george_huerta_2018,gabbard_williams_2018,schafer_nitz_2021,baltus_janquart_2021,moreno_borzyszkowski_2021,zhang_messenger_2022,andres-carcasona_mendez-vazquez_2023,gabbard_messenger_2019,chua_vallisneri_2019,khan_huerta_2020,green_gair_2020,dax_green_2021} might make it achievable if the AI can be switched from one mode to the other fast enough. Moreover, we find that even when the gap is as long as $600\,{\rm s}$, there is still an improvement in parameter estimation possible although it is smaller than for the shorter gap. We showed, further, that using the detection of the early inspiral before the long or the short gap the parameter estimation is accurate enough to not significantly limit the estimation of the frequency range that needs to be observed in resonant mode during the merger phase.

We point out, that the time before the merger as well as the criteria to select the frequencies considered were set to get a significant improvement from the detection in resonant mode. However, we did not perform a detailed search for the optimal combination and an accurate search could further improve the gain obtained by the operation of the detector in the proposed scheme. Moreover, we studied an IMBHB focusing on the detection of its merger as a representative case for AI observatories but there might be other cases where the benefit might be even bigger. Examples are the quasi-normal modes emitted during the ringdown of BH mergers that have proved difficult to detect due to their short damping time~\cite{chandrasekhar_detweiler_1975,leaver_1985,kokkotas_schmidt_1999,ligo_2021e} or spherical harmonics beyond the leading order mode that contain detailed information about the merging system~\cite{lasky_thrane_2016,calderon-bustillo_clark_2018,torres-orjuela_chen_2021}

\begin{acknowledgments}
We thank Yi-Ming Hu for helpful discussions on the detection of gravitational wave sources in multiple bands. This work was supported by the Guangdong Major Project of Basic and Applied Basic Research (Grant No. 2019B030302001) and the China Postdoctoral Science Foundation (Grant No. 2022M723676).
\end{acknowledgments}

\section*{Data Availability Statement}

The data that support the findings of this study are available from the corresponding author upon reasonable request.

\end{document}